\documentclass[11pt,3p,a4paper]{elsarticle}

\usepackage{times}

\usepackage{amsfonts, amsmath, amssymb, epsfig, indentfirst,natbib}
\usepackage[latin1]{inputenc} 
\usepackage{color}
\usepackage{placeins}
\usepackage{appendix}

\begin{document}

\title{Bias corrected minimum distance estimator for short and long memory processes}


\author[ufmg]{Gustavo C.~Lana}
\author[ufmg]{Glaura C.~Franco\corref{cor1}}
\ead{glaura@est.ufmg.br}
\author[ufmg]{Sokol~Ndreca}

\cortext[cor1]{Corresponding author}
\address[ufmg]{Departamento de Estatística, Universidade Federal de Minas Gerais, Antônio Carlos 6627, Belo Horizonte, MG, Brazil. 31270-901.}



\begin{abstract}
This work proposes a new minimum distance estimator (MDE) for the parameters
of short and long memory models. This bias corrected minimum
distance estimator (BCMDE) considers a correction in the usual MDE to
account for the bias of the sample autocorrelation function when the mean
is unknown. We prove the weak consistency of the BCMDE for the general
fractional autoregressive moving average (ARFIMA($p, d, q$)) model and derive its asymptotic distribution for some particular cases.
Simulation studies show that 
the BCMDE presents a good performance compared to other procedures frequently used in the literature, such as the maximum likelihood
estimator, the
Whittle estimator and the MDE.
The results also show that the BCMDE presents, in general, the smallest mean squared error and is less biased than the MDE when
the mean is a non-trivial function of time.
\end{abstract}


\maketitle



\newpage


\section{Introduction}

One of the most used approaches in the time series context is the autoregressive moving average (ARMA$(p,q)$) model
(Box and Jenkins (1976)). If the series presents the long memory characteristic, the ARMA procedure can be generalized to the fractional autoregressive moving average 
(ARFIMA$(p,d,q)$) model (Hosking, 1981). It can be noted that the ARFIMA model becomes an ARMA model when the memory parameter, $d$, is equal to zero. The main characteristic of the ARFIMA, which differenciates it from the ARMA model, is that its autocorrelation function decays slowly to zero, and as a result it is not absolutely summable. There are many different approaches to estimate parameters in ARMA and ARFIMA processes. In ARMA models, the most commonly used method is the maximum likelihood estimator (MLE), while in ARFIMA models the Whittle estimator (Whittle, 1951; Fox and Taqqu, 1986), which uses the periodogram function, is the preferred methodology.

Some approaches suggest the direct use of autocorrelation functions to estimate the parameters in ARFIMA models. Back in 1986, Andel suggested using the first sample autocorrelation to estimate $d$ for ARFIMA$(0,d,0)$ models. Tieslau {\it et al.} (1996) introduced the minimum distance estimator (MDE), which allows the use of more than one lag of sample autocorrelations for ARFIMA$(p,d,q)$ models and derived its asymptotic distribution for $d\in(-0.5,$ $0.25)$. The MDE minimizes the distance between the sample autocorrelations and the corresponding theoretical autocorrelations. Other works use ideas similar to the MDE for long memory models, including the ones of Mayoral (2007) and Zevallos and Palma (2013). 

When the mean of the process is unknown, a very common situation in practice, sample autocovariances and autocorrelations are biased. There are many studies in the literature about this subject. For instance, Hassani {\it et al.} (2012) discuss how this bias may affect the identification of long memory processes. Arnau and Bono (2001) and Huitema and McKean (1994) suggest alternative autocorrelation estimators with lower bias. Although these alternative estimators do tend to reduce the bias, they do not take into account the fact that the bias is not a function of the sample size alone. The values of the autocorrelation in other lags can also affect the bias.

If the sample autocorrelations are biased, the same could be expected to happen to the minimum distance estimators, as they rely on these statistics. Thus, the main objective of this work is to propose a minimum distance estimator that takes the sample autocorrelation and autocovariance bias into account, instead of trying to correct them. This can be done by minimizing the distance between the sample autocorrelations and their expectations, given a set of parameters and sample size. Based on this formulation, it is expected that this estimator reduces the bias existing in the computation of the MDE. In this sense, we call this new estimator the bias corrected minimum distance estimator (BCMDE).

We show that the BCMDE is weakly consistent and evaluate its asymptotic distribution in the case of ARFIMA(0, d, 0), pure autoregressive of order 1 (AR(1)) and pure moving average of order 1 (MA(1)) models. We have also performed a large simulation study to assess the small sample properties of the BCMDE and make a comparison with some other estimators in the literature.

This work is organized in the following way.  In Section 2 the ARMA and ARFIMA models are defined and the main parameter estimation methods are reviewed. In Section 3 we study some properties of the sample autocorrelation and autocovariance functions. In Section 4 we introduce the main contribution of this work, the bias corrected minimum distance estimator. Section 5 presents simulation results in order to compare the BCMDE with other estimators in the literature. Finally, Section 6 includes conclusion and future remarks. All proofs are given in the Appendix.

\section{ARFIMA and ARMA processes}

This section presents the basic properties of ARFIMA and ARMA processes, including the autocorrelation, autocovariance and spectral function, as well as methods of estimation.

\subsection{Definition of ARFIMA and ARMA Processes}

A stochastic process $\{X_t\}$ is an ARFIMA$(p,d,q)$ (Hosking, 1981; Granger and Joyeux, 1980) process if it satisfies the equation $X_t-\mu=(1-B)^{-d}U_t$, $t\in\mathbb{Z},$ where $U_t=\frac{\theta(B)}{\phi(B)}a_t$, $\phi(B)=1-\phi_1B-...-\phi_pB^p$, $\theta(B)=1+\theta_1B+...+\theta_qB^q$, $B$ is the backward shift operator such that $B^kX_t=X_{t-k}$, $\phi_1,..., \phi_p,\theta_1,...,\theta_q$ are real numbers, $\mu$ is the mean of the process and $a_t$ is a zero-mean white noise process with $Var(a_t)=\sigma^2<\infty$.

The quantity $d$ is called the memory parameter and it can hold non-integer values. The ARFIMA process is stationary if $d<0.5$ and if all roots of $\phi(B)$ lay outside the unit circle in the complex plane. The process is called invertible if $d>-0.5$ and if $\theta(B)$ has all its roots outside the unit circle of the complex plane. If $d\in(0,0.5)$ the process has the property of long memory, characterized by an autocorrelation function that is not absolutely summable. For any ARFIMA process with $d\neq 0$, the autocorrelation and autocovariance functions decay asymptotically proportionally to $c^{2d-1}$, $c=1,2,...$.

When $d=0$ we have the short memory ARMA $(p,q)$ process, $U_t$.
A common characteristic of the autocorrelation function of ARMA processes is that it is absolutely summable.
The spectral density of the ARMA process is given by
\begin{equation} \label{eqn:EqfU}
f_U(\omega)=\frac{\sigma^2}{2\pi}\frac{|\theta(e^{-i\omega})|^2}{|\phi(e^{-i\omega})|^2}, \quad\omega\in\mathbb{R}.
\end{equation}

When $p,q=0$ the ARFIMA$(0,d,0)$ process is called a fractional white noise. In this case, for $d<0.5$, $Var(X_t)=\gamma_0=(-2d)!/(-d)!^2$. Its autocorrelation function, for $d<0.5$, is given by
\begin{equation} \label{eqn:nfacr}
\rho_k=\prod_{i=1}^k\frac{i-1+d}{i-d},\quad k\in\mathbb{Z}.
\end{equation}

If $p>0$ or $q>0$ the autocorrelation function is more difficult to be obtained, but it can be accurately calculated through the splitting method (Brockwell and Davis, 1991; Bertelli and Caporin, 2002). Following this method, if $\gamma^{(1)}_k$ is the autocovariance function of the ARMA component and $\gamma^{(2)}_k$ the autocovariance function of the fractional white noise component, then the autocovariance of the ARFIMA process, $\gamma_k$, can be decomposed as $\gamma_k=\sigma^{-2}\sum_{i=-\infty}^\infty\gamma_i^{(1)}\gamma_{i-k}^{(2)},\quad k\in\mathbb{Z}$,
and the autocorrelation function of the ARFIMA process can be calculated as
\begin{equation} \label{eqn:Eq11.Cr}
\rho_k=\frac{\sum_{i=-\infty}^\infty\gamma_i^{(1)}\gamma_{i-k}^{(2)}}{\sum_{i=-\infty}^\infty\gamma_i^{(1)}\gamma_{i}^{(2)}},\quad k\in\mathbb{Z}.
\end{equation}
The splitting method is valid even if the infinite autoregressive representation of ARFIMA models, for $d<0$, are not absolutely summable.

Let $\varrho$ be a vector of autocorrelations of an ARFIMA process with $d\in(-0.5,0.25)$ and $\hat{\varrho}$ the vector of sample autocorrelations of $\varrho$. Hosking (1996) shows that \linebreak $\sqrt{T}(\hat{\varrho}-\varrho)\xrightarrow{\mathcal{D}}N(0,C)$, where $\xrightarrow{\mathcal{D}}$ stands for convergence in distribution and $C$ is a matrix whose element $C_{ij}$ is given by
\begin{equation} \label{eqn:Eq11}
C_{ij}=\sum_{l=1}^\infty(\rho_{l-i}+\rho_{l+i}-2\rho_i\rho_l)(\rho_{l-j}+\rho_{l+j}-2\rho_{j}\rho_{l}), \quad i,j \in\mathbb{Z}.
\end{equation}

The spectral density of the ARFIMA process is given by
$$f(\omega)=\left(2\sin\frac{\omega}{2}\right)^{-2d}f_U(\omega),\quad \omega\in\mathbb{R},$$
where $f_U(\omega)$ is the spectral density of the ARMA process given in Equation$~\eqref{eqn:EqfU}$. For $d>0$, the spectral function satisfies $\lim_{\omega\to 0}f(\omega)=\infty$.

The most common estimator of the spectral function is the periodogram. Suppose  $x_{1},...,x_{T}$ is a partial realization of $\{X_{t}\}$. Hence, the periodogram function is defined as
$I_{x}(\omega) = {(2\pi T)}^{-1} |\sum_{t=1}^{T} x_{t}e^{i\omega
t}|^{2}$.

\subsection{Estimation for ARMA$(p,q)$ and ARFIMA$(p,d,q)$ models}

Let the parameter vector be defined as $\lambda=(\delta,d)'$, where $\delta=(\phi_1,...,\phi_p,\theta_1,...,\theta_q)$. So, for the ARMA model, $\lambda=\delta$.

Assuming that $a_t$ is a Gaussian white noite process, the maximum likelihood estimator (MLE) for both ARFIMA and ARMA models, is the parameter vector $\lambda$ that maximizes:
$$l(\lambda,\sigma^2)=-\frac{1}{2T}log(\sigma^{-2}\det{\Sigma_\lambda})-\frac{1}{2T\sigma^2}X'(\Sigma_\lambda)^{-1}X,$$
where $X=(X_1,...,X_T)$.

In order to evaluate the likelihood function, the necessity of calculating the determinant and the inverse of a $T\times T$ matrix can make this procedure rather unpractical. Fortunately, for ARMA processes this is not necessary, as to
maximize $l(\delta,\sigma^2)$ is the same as to maximize
$$l(\delta,\sigma^2)^*=(v_1...v_T)^{-1/2}\exp\left\{-\frac{1}{2}\sum_{t=1}^T[X_t-\hat{X}_t(\delta,t-1,...,1)]^2/v_t\right\},$$
where $v_t$, $t=1,...,T$ are the mean squared errors of $X_t-\hat{X}_t(\delta,t-1,...,1)$ and $\hat{X}_t(\delta,t-1,...,1)$ is the best predictor of $X_t$ given the value of $\delta$ and $X_1,...,X_{t-1}$. See Brockwell and Davis (1991) to learn in details how $v_1,...,v_T$ and $\hat{X}_t(\delta,t-1,...,1)$ are calculated.

In order to calculate the asymptotic covariance of the MLE, define the autoregressive processes $\mathcal{U}_t$ and $\mathcal{V}_t$ that satisfy the equation $\phi(B)\mathcal{U}_t=a_t$, $\theta(B)\mathcal{V}_t=a_t$. Here again $a_t$ is a white noise process. Finally define the vectors $\mathcal{U}=(\mathcal{U}_t,...,\mathcal{U}_{t-1+p})'$ and $\mathcal{V}=(\mathcal{V}_t,...,\mathcal{V}_{t-1+q})'$. The matrix of asymptotic covariances of the MLE estimator is given by (Brockwell and Davis, 1991)
$$Var(\hat{\delta})=\left[ \begin{array}{cc}
E(\mathcal{U}\mathcal{U}') & E(\mathcal{U}\mathcal{V}') \\
E(\mathcal{V}\mathcal{U}') & E(\mathcal{V}\mathcal{V}') \\
\end{array} \right]^{-1}.
$$

Concerning the ARFIMA process, the most common estimator for $\lambda$ is the Whittle estimator (Whittle, 1951; Fox and Taqqu, 1986). The Whittle estimator is based on minimizing an approximation of the log-likelihood function, given by
\begin{equation} \label{eqn:Eq12}
l_w(\lambda,\sigma^2)=\sum_{j=1}^{\lfloor T/2\rfloor}\left[\log f_{\lambda,\sigma^2}(\omega_j)+\frac{I(\omega_j)}{f_{\lambda,\sigma^2}(\omega_j)}\right],
\end{equation}
where $\omega_j=2\pi j/T$, $j=1,2,...,\lfloor T/2\rfloor$, are the Fourier frequencies, and $f_{\lambda,\sigma^2}$ is the spectral function given $\lambda$ and $\sigma^2$. Numerical procedures are necessary to find the values of $\lambda$ and $\sigma^2$ that minimize$~\eqref{eqn:Eq12}$.

The asymptotic distribution of the Whittle estimator $\hat{\lambda}_w$ is given by \linebreak$\sqrt{T}(\hat{\lambda}_w-\lambda)\xrightarrow{\mathcal{D}}N(0,V^{-1})$, where $V$ is a matrix with elements $V_{ij}$ given by
$$
V_{ij}=\frac{1}{4\pi}\int_{-\pi}^\pi \left[\frac{\partial\log f_{\lambda,\sigma^2}(\omega)}{\partial\lambda_i}\right]\left[\frac{\partial\log f_{\lambda,\sigma^2}(\omega)}{\partial\lambda_j}\right]d\omega.
$$
The Whittle estimator combines relatively little computational complexity and good accuracy (Palma, 2007; Rea {\it et al.}, 2013).

The minimum distance estimator (MDE), $\hat{\lambda}_{\textrm{mde}}$, for ARFIMA processes was proposed by Tieslau {\it et al.} (1996). The idea is to minimize the difference between the theoretical autocorrelations and the sample autocorrelations. Although the MDE was proposed as an estimator for ARFIMA processes, it can also be used in ARMA models. Define $\hat{\varrho}$ as a vector of sample autocorrelations and $\varrho(\lambda)$ as the vector of corresponding theoretical autocorrelations, given the parameter vector $\lambda$. The minimum distance estimator is the one that minimizes
$S(\lambda)=[\hat{\varrho}-\varrho(\lambda)]'W[\hat{\varrho}-\varrho(\lambda)]$,
where $W$, the weighting matrix, is a symmetric, positive-definite matrix. The asymptotically optimal $W$ matrix is $W=C^{-1}$, where $C$ is the asymptotic
covariance matrix of the sample autocorrelations (Tieslau {\it et al.}, 1996) whose elements are given in Equation$~\eqref{eqn:Eq11}$. It should be noted, though, that if the parameters are unknown, so is $C$. Tieslau {\it et al.} (1996) show that for $d\in(-0.5,0.25)$, $\sqrt{T}(\hat{\lambda}_{\textrm{mde}}-\lambda)\xrightarrow{\mathcal{D}} N(0,(D'WD)^{-1}D'WCWD(D'WD)^{-1})$ when $T\to\infty$, where $D$ is the matrix of derivatives of $\rho(\lambda)$ with respect to $\lambda$.


As it can be seen from the definition of the MDE, it depends heavily on estimators of the autocorrelation function, $\varrho(\lambda)$. If we use biased estimators for $\varrho(\lambda)$, this is likely to cause a negative impact on the MDE. 

In the next section we present the most used estimators of the autocovariance and autocorrelation functions and show that they are biased, implying that a correction will be needed in order to obtain better minimum distance estimators.

\section{Bias in the sample autocovariance and autocorrelation functions}

Estimation of the autocovariance, autocorrelation and spectral functions are important not only because they help to identify the correct model, but also because they are frequently used in the estimation of model parameters. Therefore, it is essential to understand their behavior and how an under (or over) estimation of these functions may affect parameter estimation.

Let $\{X_t\}$, $t\in\mathbb{Z}$, be a process with $E(X_t)=\mu_t$ and such that $X_t-\mu_t$ is stationary.
For a realization of size $T$ of $\{X_t\}$, one possible estimator for the autocovariance, $\gamma_k$, is given by,
\begin{equation} \label{eqn:Eq2}
\hat{\gamma}_k=\frac{\sum_{j=1}^{T-k}(X_j-\hat{\mu}_j)(X_{j+k}-\hat{\mu}_{j+k})}{T-k},\quad k=0,...,T-1,
\end{equation}
where $\hat{\mu}_j$ and $\hat{\mu}_{j+k}$ are estimates of $\mu_j$ and $\mu_{j+k}$. In the case in which the mean of $X_t$ is constant, then $\hat{\mu}_j=\hat{\mu}_{j+k}=\bar{X}$. Using $T$ instead of $T-k$ in Equation$~\eqref{eqn:Eq2}$ is more common in the literature, but here we will employ $\hat{\gamma}_k$ to calculate the bias corrected minimum distance estimator, which is defined in Section 4.

\subsection{Calculating the bias}

If the mean of the process is known, we can use its true value in Equation$~\eqref{eqn:Eq2}$, instead of an estimate. Thus, the following estimator can be built:

$$\tilde{\gamma}_k=\frac{\sum_{j=1}^{T-k}(X_j-\mu_j)(X_{j+k}-\mu_{j+k})}{T-k}, \quad k=0,...,T-1.$$

It is easy to verify that $E(\tilde{\gamma}_k)=\gamma_k$. This unbiased estimator under known mean, $\tilde{\gamma}_k$, has some drawbacks. For example, the variance of this estimator when $k$ is big becomes excessively large.
This can result in an estimation of the spectral function that is very inaccurate.
The sample autocovariance is also impaired by the high variance  at higher lags, with the autocovariance function assuming unusual values at these lags. This happens because when $k$ is close to $T$, $\tilde{\gamma}_k$ is based on fewer sums of the quantity $(X_j-\mu_j)(X_{j+k}-\mu_{j+k})$. For minimum distance estimators, though, this is not an issue. The MDF tends to use the smaller lags of the autocovariance function, so the large variance of $\tilde{\gamma}_k$ at higher lags will not cause any damage on them.

In practice, though, a known mean is a rare situation. Thus, the estimation of the autocovariance function is usually performed using $\hat{\gamma}_k$. The issues regarding the variance at higher lags discussed in the previous paragraph are also present in these estimators. Furthermore, $\hat{\gamma}_k$ is a biased estimator for the autocovariance function, a fact which is already proven in the literature (see Priestley, 1981).

The following proposition establishes the expectation of $\hat{\gamma}_k$ when the mean is constant as an equation with number of operations of order $T$.

\newtheorem{pro1}{Proposition}
\begin{pro1}
The expectation of $\hat{\gamma}_k$ when the mean is constant is given by
\begin{equation} \label{eqn:Prop1}
E(\hat{\gamma}_k)=\gamma_k-\frac{T+k}{T-k}\left[\frac{T\gamma_0+\sum_{i=1}^{T-1}2(T-i)\gamma_i}{T^2}\right]+2\frac{\sum_{i=1}^k\sum_{j=1}^T\gamma_{|i-j|}}{T(T-k)}.
\end{equation}
\textit{Proof in Appendix A.} 
\end{pro1}


The bias of $\hat{\gamma}_k$ originates from the fact that $\bar{X}$ is used to estimate $\mu$. We can note that the term
$[T\gamma_0+\sum_{i=1}^{T-1}2(T-i)\gamma_i]/T^2$ in  Equation$~\eqref{eqn:Prop1}$ is the variance of $\bar{X}_t$. Indeed, Priestly (1981) had already shown that $E(\hat{\gamma}_k)\approx\gamma_k-Var(\bar{X})$. For any estimator of the autocovariance function, the autocorrelation at lag $k$ is given by $\rho_k=\gamma_k/\gamma_0$.

Define now the functions $B_{T,k}^\gamma=\gamma_k-E(\hat{\gamma}_k)$, which gives the bias of $\hat{\gamma}_k$ for samples of size $T$ and $B_{T,k}^\rho=B_{T,k}^\gamma/\gamma_0$. In the particular case in which the mean of the process is a constant and is estimated as $\hat{\mu}=\sum_{t=1}^TX_t/T$, we have that
\begin{equation} \label{eqn:Eq4}
B_{T,k}^\gamma=-\frac{T+k}{T-k}\left[\frac{T\gamma_0+\sum_{i=1}^{T-1}2(T-i)\gamma_i}{T^2}\right]+2\frac{\sum_{i=1}^k\sum_{j=1}^T\gamma_{|i-j|}}{T(T-k)},
\end{equation}
and
\begin{equation} \label{eqn:Eq5}
B_{T,k}^\rho=-\frac{T+k}{T-k}\left[\frac{T\rho_0+\sum_{i=1}^{T-1}2(T-i)\rho_i}{T^2}\right]+2\frac{\sum_{i=1}^k\sum_{j=1}^T\rho_{|i-j|}}{T(T-k)}.
\end{equation}
Note that both $B_{T,k}^\gamma$ and $B_{T,k}^\rho$ are weighed means of $\gamma_0,...,\gamma_{T-1}$. For instance, we can write $B_{T,k}^\gamma$ as $B_{T,k}^\gamma=\sum_{i=0}^{T-1}w_{T,k,i}\rho_i$. In the case in which $k=1$, we have that
$$w_{T,1,i}=-\frac{T+1}{T-1}\left[\frac{T}{T^2}\right]+\frac{2}{T(T-1)},\quad i=0$$
$$w_{T,1,i}=-\frac{T+1}{T-1}\left[\frac{2(T-i)}{T^2}\right]+\frac{2}{T(T-1)},\quad i=1,...,T-1.$$

\subsection{Empirical analysis of the bias in the sample autocovariance and autocorrelation functions for a constant mean}

In Subsection 3.1 we have derived the exact values of the bias of the sample autocovariance function as a function of the autocovariance at lags $0,...,T-1$. Calculation of the bias for the sample autocorrelation is far more complicated. In this subsection we provide some numerical examples of how the sample autocovariance function can be affected by the bias in its estimation through a comparison with the theoretical function. In the case of the autocorrelation function, we discuss an approximation for calculating the expectation of its estimator.

Figure 1 shows, for an ARFIMA$(0,0.3,0)$ process with $T=100$, the difference in the behavior of the expectation of the autocovariance estimator $\hat{\gamma}_k$ compared to the theoretical autocovariance function. By analyzing Figure 1, some interesting conclusions can be reached. The first one is that, for all lags, the sample autocovariance is far from its expected value. The second one is that the expectation of the sample autocovariance appears to behave more like the autocovariance functions of short memory process, possessing a fast decay.

\footnotesize{
\begin{figure}
\caption{Autocovariance function (bars) and expectation (full line) of the estimator $\hat{\gamma}_k$ for ARFIMA(0,0.3,0) with $T=100$.}
\includegraphics[width=\textwidth]{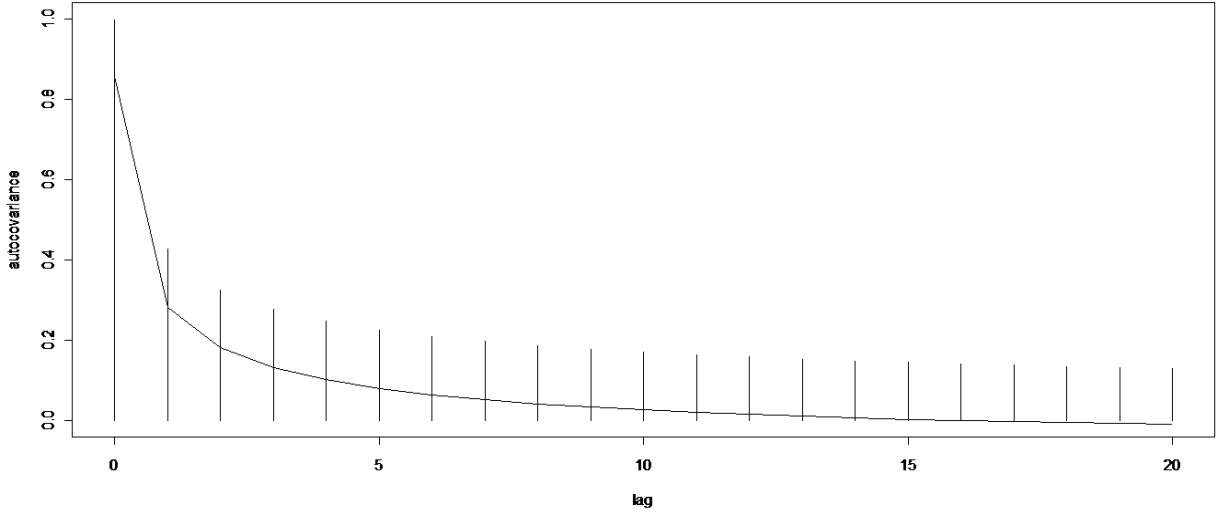}
\end{figure}
}

\normalsize

For any estimator of the autocovariance function, the autocorrelation at lag $k$ can be estimated as $\hat{\rho}_k=\hat{\gamma}_k/\hat{\gamma}_0$. A derivation of the bias for the sample autocorrelations is more complicated, and it is likely distribution-dependent. In this work we suggest to approximate $E(\hat{\rho}_k)$ by $E(\hat{\gamma}_k)/E(\hat{\gamma}_0)$. This approach makes sense asymptotically, as $E(\hat{\gamma}_k)\to\gamma_k$ and $E(\hat{\gamma}_0)\to\gamma_0$ and therefore $E(\hat{\gamma}_k)/E(\hat{\gamma}_0)\to\rho_k$. To verify if it works  reasonably well also for small samples, a Monte Carlo study with 1000 replications was performed for an AR$(1)$ process with $T=50$. Table 1 shows the results of this study for $\phi=0.4,0.6,0.8$ and $k=1,2,3$. In this simulation we compared the approximation $E(\hat{\gamma}_k)/E(\hat{\gamma}_0)$ with $\widehat{E(\hat{\rho}_k)}$, which is the mean of the sample autocorrelations in the 1000 replications (an empirical estimation of $E(\hat{\rho}_k)$). From the results it seems the approximation is adequate, although it tends to overestimate $\widehat{E(\hat{\rho}_k)}$.

\footnotesize{
\begin{table}
\footnotesize{
\caption{Monte Carlo simulation for AR$(1)$ with $\phi=0.4,0.6,0.8$ and $T=50$. $E(\hat{\gamma}_k)/E(\hat{\gamma}_0)$ are the ratios of the means of autocovariance estimators, and $\widehat{E(\rho_k)}$ is the estimated mean of the autocorrelation estimator.}
\begin{center}
\begin{tabular}{ccccc}
\hline
\hline
 & & $\rho_1$ & $\rho_2$ & $\rho_3$ \\
\hline
$\phi=0.4$ & $E(\hat{\gamma}_k)/E(\hat{\gamma}_0)$ & 0.3707 & 0.1192 & 0.0186 \\
 & $\widehat{E(\hat{\rho}_k)}$ & 0.3557 & 0.1050 & 0.0005 \\
\hline
$\phi=0.6$ & $E(\hat{\gamma}_k)/E(\hat{\gamma}_0)$ & 0.5654 & 0.3054 & 0.1494 \\
 & $\widehat{E(\hat{\rho}_k)}$ & 0.5378 & 0.2746 & 0.1250 \\
\hline
$\phi=0.8$ & $E(\hat{\gamma}_k)/E(\hat{\gamma}_0)$ & 0.7576 & 0.5663 & 0.4131 \\
 & $\widehat{E(\hat{\rho}_k)}$ & 0.7287 & 0.5198 & 0.3613 \\
\hline
\hline
\end{tabular}
\end{center}
}
\end{table}
}

\normalsize

With the results of Table 1 in mind, we can further investigate the approximation $E(\hat{\rho}_k)\approx E(\hat{\gamma}_k)/E(\hat{\gamma}_0)$. The exact value of $E(\hat{\rho}_k)$ is given by
$$E(\hat{\rho}_k)=E(\hat{\gamma}_k)E(\hat{\gamma}_0^{-1})+\mathrm{Cov}(\hat{\gamma}_k,\hat{\gamma}_0^{-1}).$$
The quantities $E(\hat{\gamma}_0^{-1})$ and $\mathrm{Cov}(\hat{\gamma}_k,\hat{\gamma}_0^{-1})$ are both unknown. It is possible to state, though, that $E(\hat{\gamma}_0^{-1})\geq E(\hat{\gamma}_0)^{-1}$, and therefore $E(\hat{\gamma}_k)E(\hat{\gamma}_0^{-1})\geq E(\hat{\gamma}_k)E(\hat{\gamma}_0)^{-1}$. This is because $\hat{\gamma}_0$ is a positive random variable and $f(x)=1/x$ is a convex function on $\mathbb{R}^+$ (Jensen's inequality). As in the simulations we have encountered $\widehat{E(\hat{\rho}_k)}<E(\hat{\gamma}_k)/E(\hat{\gamma}_0)$, this underestimation of the bias (and overestimation of the mean) seems to be caused by the negative autocovariance between $\hat{\gamma}_k$ and $\hat{\gamma}_0^{-1}$ in this case.

We end this subsection by making a comparison between the
theoretical autocorrelations with the approximation for the expectation of the sample autocorrelations given by $E(\hat{\gamma}_k)/E(\hat{\gamma}_0)$. This can be seen in Figure 2 for an ARFIMA$(0,0.3,0)$ process with $T=100$. We can draw similar conclusions to those made in Figure 1, that is, strong bias and behavior similar to short memory processes. Naturally there are caveats about the fact that we are dealing with approximations in Figure 2.

\footnotesize{
\begin{figure}
\caption{Autocorrelation function (bars) and approximation for the expectation (full line) of the estimator $\hat{\rho}_k$ for ARFIMA(0,0.3,0) with $T=100$.}
\includegraphics[width=\textwidth]{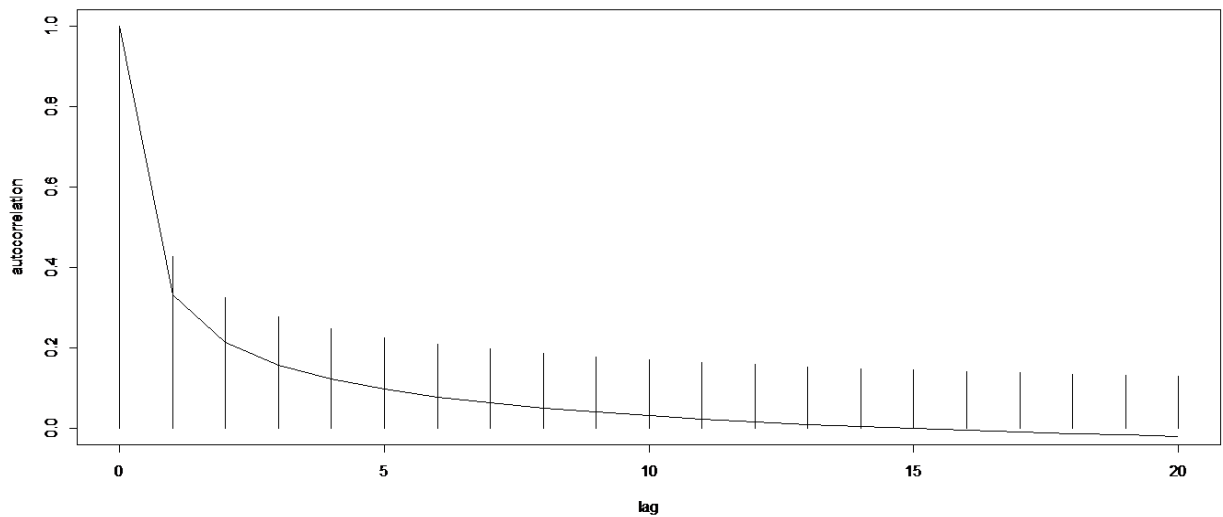}
\end{figure}
}

\normalsize

\subsection{Bias in the sample autocovariance for a particular case of a nonconstant mean}

Time series models may be generalized to cases where the mean is nonconstant in time.
In what follows, let $\mu_t$ be the mean of the process at time $t$. In the previous subsections, it was shown that the mean estimation in a model with constant mean causes bias in the autocovariance estimators. Intuitively, the same could be expected to happen if the mean is nonconstant. We will investigate the particular case in which $\mu_t=\alpha+\beta z_t$, $t=1,...,T$, where $z_1,...,z_T$ are non-stochastic regressor variables.


In the case of a simple linear regression model, given the independent variables, $z_1,...,z_T$, the parameters $\alpha$ and $\beta$ in the equation $\mu_t=\alpha+\beta z_t$ can be estimated through the ordinary least squares method.
Then the estimator of $\mu_t$ is given by
\begin{align}
\begin{split} \label{eqn:em}
\hat{\mu}_t&=\hat{\alpha}+\hat{\beta}z_t\\
&=\bar{X}+\hat{\beta}\tilde{z}_t,
\end{split}
\end{align} 
where $\tilde{z}_t=z_t-\bar{z}$.

\newtheorem{pro2}[pro1]{Proposition}
\begin{pro2}
The expectation of $\hat{\gamma}_k$ when the mean is a linear function of a univariate regressor variable is given by
\begin{align}
\begin{split} \label{prop2} 
E(\hat{\gamma}_k)&=\gamma_k-\frac{2\sum_{t=1}^{T-k}\sum_{i=1}^T\gamma_{|t-i|}}{(T-k)T}-\frac{\sum_{t=1}^{T-k}\sum_{i=1}^T(\tilde{z}_{t+k}\tilde{z}_i\gamma_{|t-i|}+\tilde{z}_t\tilde{z}_i\gamma_{|t+k-i|})}{(T-k)\sum_{i=1}^T\tilde{z}_i^2}\\
&+\frac{[\sum_{t=1}^{T-k}(\tilde{z}_t+\tilde{z}_{t+k})]\sum_{i=1}^T\sum_{j=1}^T\tilde{z}_j\gamma_{|i-j|}}{(T-k)T\sum_{i=1}^T\tilde{z}_i^2}\\
&+\frac{[\sum_{t=1}^{T-k}\tilde{z}_t\tilde{z}_{t+k}]\sum_{i=1}^T\sum_{j=1}^T\tilde{z}_i\tilde{z}_j\gamma_{|i-j|}}{(T-k)(\sum_{i=1}^T\tilde{z}_i^2)^2}+\frac{\sum_{i=1}^T\sum_{j=1}^T\gamma_{|i-j|}}{T^2}.
\end{split}
\end{align}
\end{pro2}


In the particular case where $z_t=z_{t-1}+1$, $t=2,...,T$, as when $z_t$ is the time, the formula of the expectation of $\hat{\gamma}_k$ can be simplified to
$$E(\hat{\gamma}_k)=\gamma_k-\frac{2\sum_{t=1}^{T-k}\sum_{i=1}^T\gamma_{|t-i|}}{(T-k)T}-\frac{24\sum_{t=1}^{T-k}\sum_{i=1}^T\tilde{z}_{t+k}\tilde{z}_i\gamma_{|t-i|}}{(T-k)(T^3-T)}$$
$$+\frac{12[(T-k)^3-(T-k)(3k^2+1)]\sum_{i=1}^T\sum_{j=1}^T\tilde{z}_i\tilde{z}_j\gamma_{|i-j|}}{(T-k)(T^3-T)^2}+\frac{\sum_{i=1}^T\sum_{j=1}^T\gamma_{|i-j|}}{T^2}.$$

The proof of Proposition 2 is straightforward using properties of the expectation in a similar fashion to what was done in Proposition 1. 

\section{Bias corrected minimum distance estimator}

This section focuses on the main proposal of this work, the bias corrected minimum distance estimator (BCMDE). As we have showed in the previous sections, the usually employed estimators of the autocorrelation function are biased due to the fact that the mean generally needs to be estimated. As the MDE is calculated using the autocorrelation function, in this section we propose a new minimum distance estimator that tries to take this bias into account.

The idea of the BCMDE is to minimize the distance, not between sample and theoretical autocorrelations, but between sample autocorrelation and its expectation, using the approximation $E(\hat{\rho}_k)\approx E(\hat{\gamma}_k)/E(\hat{\gamma}_0)$, which seems to be a reliable approximation for the expectation of the sample autocorrelation, as seen in Subsection 3.2.

Let $\rho_{T,k}$ be the ratio between $E(\hat{\gamma}_k)$ and $E(\hat{\gamma}_0)$, whose formula was given in Proposition 1. Thus $\rho_{T,k}$ can be written as
\begin{equation} \label{eqn:Ro}
\rho_{T,k}=\frac{\rho_k+B_{T,k}^\rho}{1+B_{T,0}^\rho}, \quad k=0,...,T-1,
\end{equation}
where $B_{T,k}^\rho$ is given in Equation$~\eqref{eqn:Eq5}$.

Let $\hat{\varrho}$ be a vector of sample autocorrelations and $\varrho_T$ be the vector corresponding to $\rho_{T,k}$. The BCMDE minimizes
\begin{equation} \label{eqn:Eq13}
S(\lambda)=(\hat{\varrho}-\varrho_T)'W(\hat{\varrho}-\varrho_T).
\end{equation}
The weighting matrix $W$ in Equation$~\eqref{eqn:Eq13}$ can be any symmetric positive definite matrix in order that $S(\lambda)$ becomes a measure of distance (not in the strict mathematical sense) between $\hat{\varrho}$ and $\varrho$. It can be the matrix of asymptotic covariances of the sample autocorrelations or the identity matrix (specially in cases where the asymptotic covariances are unknown). An obvious question regarding the vector of sample autocorrelation, $\hat{\varrho}$, is which lags should be chosen to be part of $\hat{\varrho}$. For an ARFIMA or ARMA model, the most intuitive choice for the lags in the vectors $\hat{\varrho}$ and $\varrho$ is $1,...,m$, $m\in\mathbb{N}$. Tieslau {\it et al.} (1996) showed that for an ARFIMA$(0,d,0)$ process this choice reduces the asymptotic variance of the MDE estimator compared to the choice of the lags $k,...,k+m$, for any $k\geq 2$.

We will proceed now to state the weak consistency of the BCMDE. For the proof to be valid, it is necessary first to establish the following lemma and Proposition 3.

\newtheorem{lem1}{Lemma}
\begin{lem1}
Consider the function,
$$f_T=\sum_{i=0}^{\infty}w_{T,i}a_i,\quad T\in\mathbb{N},$$
where $a_i$ and $w_{T,i}$ are sequences of real numbers satisfying the following conditions:
\begin{enumerate}[i.]
\item $a_i\to 0$ as $i\to\infty$.
\item There exists a real number $U>0$ such that $\sum_{i=1}^\infty|w_{T,i}|\leq U$ for a sufficiently large $T$.
\item For a fixed $i$, $w_{T,i}\to 0$ as $T\to\infty$.
\end{enumerate}
Then the function $f_T$ goes to zero as $T\to\infty$.

\end{lem1}

\newproof{prolem1}{Proof}
\begin{prolem1}
Let $\epsilon>0$. Define $i_0$ such that $a_i<\epsilon$ for $i>i_0$. Note that $f_T=\sum_{i=0}^{i_0}w_{T,i}a_i+\sum_{i=i_0+1}^\infty w_{T,i}a_i$. The sum $\sum_{i=0}^{i_0}w_{T,i}a_i$ goes to zero as $T\to\infty$ due to the fact that $w_{T,i}\to 0$. The second sum is bounded by $\epsilon U$.
$\square$
\end{prolem1}

\newtheorem{pro3}[pro1]{Proposition}
\begin{pro3}
Suppose that the autoregressive coefficients $\phi_1,...,\phi_p$ of an ARFIMA$(p,d,q)$ satisfy $\sum_{i=1}^p|\phi_i|\leq K<1$. Then the autocorrelation function of an ARFIMA$(p,d,q)$ converges uniformly to zero as $T\to\infty$ in a compact parametric space. \textit{Proof in Appendix B.}
\end{pro3}


Note that Proposition 3 is also valid for an ARMA$(p,q)$ process, which is a special case of the ARFIMA process.

With the help of Proposition 3, Theorem 1 below provides the weak convergence of the BCMDE.

\newtheorem{the1}{Theorem}
\begin{the1}
Let $\{X_t\}$ be a stationary linear process and assume that the function $B_{T,k}^\rho$ converges uniformly to $0$ as $T\to\infty$ in a compact parametric space $\Lambda$. In addition, let the vector of theoretical autocovariances $\varrho=(\rho_{k_1},...,\rho_{k_m})$, $\{k_1,...,k_m\}\subset\mathbb{N}^m$, be such that $\varrho:\Lambda\to\mathbb{R}^m$ is injective. Then, $\hat{\lambda}$ converges in probability to $\lambda_0$, the real parameter values, as $T\to\infty$. \textit{Proof in Appendix C.}
\end{the1}

In some cases it is easy to check and guarantee that $\varrho$ is injective. For example, in the case of ARFIMA$(0,d,0)$ processes, the first theoretical autocorrelation is monotonous as a function of $d$. Therefore, if the first lag of the autocorrelation is present, $\varrho$ is injective. It is even more trivial to guarantee the injectivity of $\varrho$ in the case of AR(1) or MA(1) processes.

The following propositions show that the conditions imposed on $B_{T,k}^\rho$ in Theorem 1 are satisfied for the cases of constant mean and mean as linear function of time.

\newtheorem{pro4}[pro1]{Proposition}
\begin{pro4}
In the case of a model with constant mean and  $\hat{\mu}=\bar{X}$, $B_{T,k}^\rho$ converges uniformly to $0$ as $T\to\infty$ in a compact parametric space if the autocorrelation function also does so.
\end{pro4}
\newproof{propro4}{Proof}
\begin{propro4}
Consider $w_{T,k,i}$ as the weight of the $i$-th autocorrelation of $B_{T,k}^\rho(\lambda)$ (given in Equation$~\eqref{eqn:Eq5}$). We can determine an upper bound for the sum of the weights, given by $\sum_{i=0}^{T-1}|w_{T,k,i}|\leq\frac{T+k}{T-k}+\frac{Tk}{T(T-k)},$
which converges to one as $T\to\infty$ when $k$ is fixed.
\end{propro4}



Note also that, for any $i$,
\begin{equation} \label{eqn:wei}
|w_{T,k,i}|\leq\frac{(T+k)}{(T-k)}\frac{2(T-1)}{T^2}+\frac{4k}{(T-k)},
\end{equation}
which goes to zero as $T\to\infty$.

Using Lemma 1 and the uniform convergence of the autocorrelation function, the proof is completed. $\square$

\newtheorem{pro5}[pro1]{Proposition}
\begin{pro5}
In the case of a model in which the mean is a linear function of time, $\mu_t=\alpha+\beta t$, 
then $B_{T,k}^\rho$ converges uniformly in a compact parametric space if the autocorrelation function also does so. \textit{Proof in Appendix D.}
\end{pro5}

then $B_{T,k}^\rho$ converges uniformly in a compact parametric space if the autocorrelation function also does so. \textit{Proof in Appendix D}

The following result establishes the asymptotic distribution of the BCMDE in the case of an ARFIMA$(0,d,0)$ process.

\newtheorem{the2}[the1]{Theorem}
\begin{the2}
Let $\varrho=(\rho_{k_1},...,\rho_{k_m})$, $\{k_1,...,k_m\}\subset\mathbb{N}^m$, be such that $\varrho:\Lambda\to\mathbb{R}$ is injective and assume that the function $B_{T,k}^\rho$ converges uniformly in a compact parametric space $\Lambda$. If $\{X_t\}$ is an ARFIMA$(0,d,0)$ with $d<0.25$ and constant mean, then as $T\to\infty$,
$$\sqrt{T}(\hat{\lambda}-\lambda)\to N(0,(D'WD)^{-1}D'WCWD(D'WD)^{-1}),$$
where $D$ is the matrix of derivatives of $\varrho$ with respect to the parameters. \textit{Proof in Appendix E.}
\end{the2}

\newdefinition{rem}{Remark}
\begin{rem}
Theorem 2 is also valid for AR$(1)$ and MA$(1)$ processes. Nevertheless, it is worth mentioning some particularities. The bias of the sample autocovariance of both the AR$(1)$ and the MA$(1)$ processes decay faster than $T^{-1/2}$. The derivative of the autocorrelation function of the AR$(1)$ process is given by $\rho_k'(\phi)=k\phi^{k-1}$. In a compact parametric space, $\rho_k'(\phi)$ clearly converges uniformly to zero. The derivative of the autocorrelation function of the MA$(1)$ process is zero after the first lag. Finally, as in the case of the ARFIMA$(0,d,0)$ process, a non-zero derivative is guaranteed adding the first lag to $\varrho$.
\end{rem}

The simulations of next section show that, for finite sample sizes, the BCMDE has a similar performance to the MLE in the case of constant mean, and it is superior to the other estimators in the case of nonconstant mean.

\section{Monte Carlo simulations}

In order to investigate the small sample properties of the BCMDE and compare it to the other estimators described in Subsection 2.2, an extensive Monte Carlo study was performed. For this purpose, 1000 Monte Carlo replications were used with a burn-in of size 500 for each series.

In these simulations we tested short memory models (AR$(1)$ and MA$(1)$) and long memory models (ARFIMA$(0,d,0)$). In the case of short memory models we compared the BCMDE with the Whittle estimator, the MLE and the MDE. In the case of long memory models, we compared the BCMDE with the Whittle estimator and the MDE.

In all simulations the errors were generated from a standard normal distribution. For the BCMDE, only the first sample autocorrelation was used. For each estimator, mean, standard deviation (SD) and square root of the mean squared error  (RMSE) were calculated and used to compare the results.

\subsection{AR$(1)$ and MA$(1)$ models with constant mean}


In this subsection we compare the performance of the BCMDE with its competing estimators for AR$(1)$ and MA$(1)$ models with constant and unknown mean. The parameter values 
were fixed at $0.4$ and $0.8$ for both $\phi$ and $\theta$. The sample sizes used were $T=25$ and $T=100$.

The results are presented in Table 2. For $\phi=0.8$, the BCMDE has a significantly better performance not only in terms of bias, but also in terms of the RMSE. For $\phi=0.8$, the Whittle estimator presented smaller RMSE, but the BCMDE had smaller bias. Contrary to the autoregressive case, in the moving average models the BCMDE, as well as the MDE, do not present a satisfactory performance.

The contrast between the performance of the estimators in the MA(1) and AR(1) cases deserves some comments. The bias of the autocovariance estimators are linear combinations of the autocovariance function. In the MA(1) model, the autocovariance function is zero for lags greater than one, causing the bias of sample autocovariances to be irrelevant.

\footnotesize{
\begin{table}
\footnotesize{
\caption{Mean, standard deviation and RMSE for AR$(1)$ and MA$(1)$, fixed mean.}
\begin{center}
\begin{tabular}{c|cccc|cccc}
\hline
\hline
& \multicolumn{4}{c|}{AR$(1)$}  & \multicolumn{4}{c}{MA$(1)$} \\
\hline
\hline
 & Whittle & MLE & MDE & BCMDE & Whittle & MLE & MDE & BCMDE \\
\hline
& \multicolumn{4}{c|}{$T=25, \quad \phi=0.4$ }  & \multicolumn{4}{c}{$T=25,\quad\theta=0.4$} \\
\hline
Mean & 0.3551 & 0.3124 & 0.3090 & \textbf{0.3699} & \textbf{0.3868} & 0.3786 & 0.3676 & 0.4422 \\
SD & 0.1947 & 0.1884 & \textbf{0.1862} & 0.1978 & \textbf{0.2310} & 0.2796 & 0.2973 & 0.3111 \\
RMSE & \textbf{0.1997} & 0.2077 & 0.2072 & 0.2000 & \textbf{0.2313} & 0.2803 & 0.3111 & 0.3137 \\
\hline
& \multicolumn{4}{c|}{$T=100, \quad \phi=0.4$ }  & \multicolumn{4}{c}{$T=100,\quad\theta=0.4$} \\
\hline
Mean & 0.3876 & 0.3774 & 0.3775 & \textbf{0.3918} & 0.3956 & 0.3936 & \textbf{0.3992} & 0.4194 \\
SD & 0.0960 & \textbf{0.0950} & 0.0955 & 0.0966 & 0.0991 & \textbf{0.0990} & 0.1590 & 0.1660 \\
RMSE & \textbf{0.0967} & 0.0976 & 0.0981 & 0.0969 & 0.0992 & \textbf{0.0991} & 0.1589 & 0.1671 \\
\hline
\hline
& \multicolumn{4}{c|}{$T=25, \quad \phi=0.8$ }  & \multicolumn{4}{c}{$T=25,\quad\theta=0.8$} \\
\hline
Mean & 0.7208 & 0.6616 & 0.6432 & \textbf{0.7361} & 0.7212 & \textbf{0.8135} & 0.6285 & 0.7039 \\
SD & 0.2024 & \textbf{0.1602} & 0.1646 & 0.1852 & \textbf{0.1843} & 0.1870 & 0.3106 & 0.2938 \\
RMSE & 0.2172 & 0.2117 & 0.2273 & \textbf{0.1958} & 0.2004 & \textbf{0.1874} & 0.3547 & 0.3090 \\
\hline
& \multicolumn{4}{c|}{$T=100, \quad \phi=0.8$ }  & \multicolumn{4}{c}{$T=100,\quad\theta=0.8$} \\
\hline
Mean & 0.7789 & 0.7673 & 0.7669 & \textbf{0.7864} & 0.7791 & 0.8080 & 0.7686 & \textbf{0.7929} \\
SD & 0.0688 & \textbf{0.0661} & 0.0668 & 0.0681 & 0.0803 & \textbf{0.0712} & 0.2280 & 0.2201 \\
RMSE & 0.0720 & 0.0737 & 0.0745 & \textbf{0.0694} & 0.0829 & \textbf{0.0716} & 0.2300 & 0.2201 \\
\hline
\hline
\end{tabular}
\end{center}
Obs.: In bold are the means closest to the real value of the parameter and the smallest SD and RMSE.
}
\end{table}
}

\normalsize

\subsection{AR models when the mean is a linear function of time}

In this subsection we compare the performance of the BCMDE with its competing estimators for AR$(1)$ when the mean is a linear function of time ($\mu_T=\alpha+\beta t$). The parameter values used were $\phi=0.5$ and $\phi=0.7$ and the sample sizes were $T=25$ and $T=100$.

The results are presented in Table 3. For all combinations of parameter values and sample sizes, the BCMDE presented the lowest bias and RMSE. Clearly the necessity of estimating the regression parameters heavily impact the MLE and MDE estimators, which do not try to compensate for a possible bias.

\footnotesize{
\begin{table}
\footnotesize{
\caption{Mean, standard deviation and RMSE for AR$(1)$ model with the mean as a linear function of time.}
\begin{center}
\begin{tabular}{ccccc}
\hline
\hline
 & Whittle & MLE & MDE & BCMDE \\
\hline
$\phi=0.5$, $T=50$  &  &  &  &  \\
\hline
Mean & 0.4475 & 0.4239 & 0.4226 & \textbf{0.4873} \\
SD & 0.1346 & 0.1325 & \textbf{0.1319} & 0.1401 \\
RMSE & 0.1444 & 0.1527 & 0.1529 & \textbf{0.1406} \\
\hline
$\phi=0.5$, $T=100$ &  &  &  &  \\
\hline
Mean & 0.4689 & 0.4581 & 0.4581 & \textbf{0.4891} \\
SD & 0.0933 & \textbf{0.0923} & 0.0926 & 0.0949 \\
RMSE & 0.0983 & 0.1014 & 0.1016 & \textbf{0.0955} \\
\hline
\hline
 & Whittle & MLE & MDE & BCMDE \\
\hline
$\phi=0.7$, $T=50$ &  &  &  &  \\
\hline
Mean & 0.6126 & 0.5914 & 0.5882 & \textbf{0.6667} \\
SD & 0.1324 & \textbf{0.1238} & 0.1256 & 0.1385 \\
RMSE & 0.1586 & 0.1646 & 0.1682 & \textbf{0.1424} \\
\hline
$\phi=0.7$, $T=100$ &  &  &  &  \\
\hline
Mean & 0.6617 & 0.6516 & 0.6500 & \textbf{0.6865} \\
SD & 0.0809 & \textbf{0.0795} & 0.0802 & 0.0829 \\
RMSE & 0.0895 & 0.0930 & 0.0945 & \textbf{0.0839} \\
\hline
\hline
\end{tabular}
\end{center}
Obs.: In bold are the means closest to the real value of the parameter and the smallest SD and RMSE.
}
\end{table}
}

\normalsize

\subsection{ARFIMA models when the mean is a linear function of time}

In this section we compare the performance of the Whittle, MDE and BCMDE estimators for ARFIMA models when the mean is a linear function of time. We considered the sample sizes $T=100$ and $T=500$, the values $d=0.2,0.4$ for the memory parameter . 

Table 4 shows the results for the model in which the mean is a linear function of time, that is $\mu_t=\alpha+\beta t$. We can see that the Whittle estimator is severely affected by the estimation of the mean, specially when $T=100$. The bias is smaller when $T=500$, but still bigger than the bias of the BCMDE. In general, the BCMDE has the smallest RMSE and bias.

\footnotesize{
\begin{table}
\footnotesize{
\caption{Mean, standard deviation and RMSE for ARFIMA$(0,0.3,0)$ with the mean as a linear function of time.}
\begin{center}
\begin{tabular}{c|ccc}
\hline								
\hline								
	&	Whittle	&	MDE	&	BCMDE \\
\hline
$T=100$, $d=0.2$	&		&		&	 \\
\hline
Mean	&	0.1688	&	0.1434	&	\textbf{0.1936} \\
SD	&	0.0973	&	\textbf{0.0810}	&		0.0998 \\
RMSE	&	0.1021	&	\textbf{0.0988}	&		0.1000 \\
\hline								
$T=500$, $d=0.2$	&	Whittle	&	MDE	&		BCMDE \\	
\hline
Mean	&	0.1902	&	0.1804	&		\textbf{0.1975} \\
SD	&	0.0387	&	\textbf{0.0352}	&		0.0395 \\
RMSE	&	0.0399	&	0.0402	&		\textbf{0.0395} \\
\hline								
\hline								
$T=100$, $d=0.4$	&	Whittle	&	MDE	&		BCMDE \\	
\hline
Mean	&	0.3475	&	0.2767	&		\textbf{0.3729} \\
SD	&	0.0924	&	\textbf{0.0626}	&		0.0878 \\
RMSE	&	0.1062	&	0.1383	&		\textbf{0.0919} \\
\hline								
$T=500$, $d=0.4$	&	Whittle	&	MDE	&		BCMDE \\	
\hline
Mean	&	0.3822	&	0.3349	&		\textbf{0.3936} \\
SD	&	0.0377	&	\textbf{0.0264}	&		0.0384 \\
RMSE	&	0.0417	&	0.0703	&		\textbf{0.0389} \\
\hline								
\hline								
\end{tabular}
\end{center}
Obs.: In bold are the means closest to the real value of the parameter and the smallest SD and RMSE.
}
\end{table}
}

\normalsize

\section{Concluding remarks}

This work proposes a new estimator for short and long memory models,
called here BCMDE. This estimator belongs to the class of minimum distance
estimators (MDE), which are based on the sample autocorrelation function.
Previous minimum distance estimators in the literature search for the parameter
values that minimize the distance between the sample and theoretical autocorrelations.
A problem with this approach is that
the expectation of the sample autocorrelations may differ substantially from
the theoretical autocorrelations.

We have derived the exact formula for the bias of the sample autocovariance and
have shown, empirically, that this bias, which arises when we need to estimate the
mean of the process, can not be neglected. On the other hand, the exact expectation
of the sample autocorrelation is very difficult to be obtained, but we give some
empirical evidence to show that the sample autocorrelation is also affected by
the mean estimation.

The central idea of the BCMDE is to find the parameter values that minimize
the distance between the sample autocorrelation and an approximation of its
expectation. The approximation we have chosen is the ratio of the expectation
of the sample autocovariance at lags $k$ and 0. A Monte Carlo study shows
that this is a good approximation.

We have proved the weak consistency of the BCMDE in the case of a
constant mean and we have also derived its asymptotic distribution for the
ARFIMA($0, d, 0$) ($d < 0.25$), AR(1) and MA(1) models. In these circumstances,
both the BCMDE and the MDE have the same asymptotic distribution.

For small sample sizes, simulation studies showed that the BCMDE generally outperforms its competitors in estimating the 
autoregressive parameter in the AR(1) model, both in terms of bias and RMSE, especially in the case of a nonconstat mean. Among the competitors 
was the widely used maximum likelihood estimator. In the case of an ARFIMA(0,$d$,0) model with mean as a linear function of time, the BCMDE 
has also presented better performances than the other estimators commonly used to estimate the fractional parameter $d$.

Future works could encompass the search for a better approximation for the
expectation of the sample autocorrelation and the proof of some asymptotic
properties that were not covered in this work, such as the asymptotic distribution of the estimators when more than one parameter must be estimated.


\appendix


\section{Proof of Proposition 1}

Taking the expectation in$~\eqref{eqn:Eq2}$ while adding and subtracting $\mu$ in the terms of the numerator, yields
\begin{align}
\begin{split} \label{eqn:Eq13.2}
E(\hat{\gamma}_k)&=\frac{\sum_{i=1}^{T-k}E((X_i-\mu)(X_{i+k}-\mu))}{T-k}-\frac{\sum_{i=1}^{T-k}E((X_i-\mu)(\bar{X}-\mu))}{T-k}\\
&-\frac{\sum_{i=1}^{T-k}E((X_{i+k}-\mu)(\bar{X}-\mu))}{T-k}+E((\bar{X}-\mu)^2).
\end{split}
\end{align}

The first term in the right hand side of$~\eqref{eqn:Eq13.2}$ is the autocorrelation of lag $k$, $\gamma_k$. For the second and third terms we have $E((X_i-\mu)(\bar{X}-\mu))=\frac{\sum_{j=1}^T\gamma_{|i-j|}}{T}$, while for the fourth term, $E((\bar{X}-\mu)^2)=\frac{\sum_{i=1}^T\sum_{j=1}^T\gamma_{|i-j|}}{T^2}=\frac{T\gamma_0+2\sum_{i=1}^{T-1}(T-i)\gamma_i}{T^2}$.
Thus, using these equalities on Equation$~\eqref{eqn:Eq13.2}$, yields to
\begin{align}
\begin{split} \label{eqn:Eq15}
E(\hat{\gamma}_k)&=\gamma_k-\frac{\sum_{i=1}^{T-k}\sum_{j=1}^T\gamma_{|i-j|}}{T(T-k)}-\frac{\sum_{i=1}^{T-k}\sum_{j=1}^T\gamma_{|i+k-j|}}{T(T-k)}\\
&+\frac{T\gamma_0+2\sum_{i=1}^{T-1}(T-i)\gamma_i}{T^2},
\end{split}
\end{align}
which is an equation with number of operations of order of magnitude $T^2$.

Now note that $\sum_{i=1}^T\gamma_{|i-j|}=\sum_{i=1}^T\gamma_{|i-(T-j+1)|}$. As a result $\sum_{i=1}^{T-k}\sum_{j=1}^T\gamma_{|i-j|}=\sum_{i=1}^{T-k}\sum_{j=1}^T\gamma_{|i+k-j|}$. Consequently
\begin{equation} \label{eqn:Eq15.2}
E(\hat{\gamma}_k)=\gamma_k+\frac{T\gamma_0+2\sum_{i=1}^{T-1}(T-i)\gamma_i}{T^2}-2\frac{\sum_{i=1}^{T-k}\sum_{j=1}^T\gamma_{|i-j|}}{T(T-k)}.
\end{equation}
Finally, the numerator in the last term of Equation$~\eqref{eqn:Eq15.2}$ can be written as \linebreak
$\sum_{i=1}^{T-k}\sum_{j=1}^T\gamma_{|i-j|}=\sum_{i=1}^T\sum_{j=1}^T\gamma_{|i-j|}-\sum_{i=1}^k\sum_{j=1}^T\gamma_{|i-j|}$. Therefore,
\begin{align*}
E(\hat{\gamma}_k)&=\gamma_k+\left(\frac{1}{T^2}-\frac{2}{T(T-k)}\right)\left(T\gamma_0+2\sum_{i=1}^{T-1}(T-i)\gamma_i\right)\\
&+2\left[\frac{\sum_{i=1}^k\sum_{j=1}^T\gamma_{|i-j|}}{T(T-k)}\right]\\
&=\gamma_k-\frac{(T+k)}{(T-k)}\left[\frac{T\gamma_0+\sum_{i=1}^{T-1}2(T-i)\gamma_i}{T^2}\right]+2\left[\frac{\sum_{i=1}^k\sum_{j=1}^T\gamma_{|i-j|}}{T(T-k)}\right],
\end{align*}
which is an equation with number of operations of order of magnitude $T$. $\square$

\section{Proof of Proposition 3}

In order to proof Proposition 3 we will employ the splitting method.

For a pure MA($q$) model, the autocovariance function is given by,
$$\gamma_k^{(ma)}=\sigma^2\sum_{j=0}^q\theta_j\theta_{k+j},\quad k=0,...,q,$$
where $\theta_0=1$. For $k>q$, $\gamma_k^{(ma)}=0$. Clearly, such autocovariance function converges uniformly to zero as $k\to\infty$.

For a pure AR($p$) model, the autocovariance function satisfies the recursive relation $\gamma_k=\phi_1\gamma_{k-1}+...+\phi_p\gamma_{k-p}$ for $k>0$. Thus, for any values of $\phi_1,...,\phi_p$ in the parametric space, the following generic inequality can be established: $|\gamma_k^{(\mathrm{ar})}|\leq K^{\lceil k/p\rceil}\gamma_0^{(\mathrm{ar})}$, where  $\lceil . \rceil$ is the ceiling function.

We can also determine a bound for the sum of the autocovariance function, a result that will be important in the remaining of the proof,
$$
\bigg|\sum_{k=0}^\infty\gamma_k^{(\mathrm{ar})}\bigg|\leq\gamma_0^{(\mathrm{ar})}+\gamma_0^{(\mathrm{ar})}\sum_{k=1}^\infty K^{\lceil k/p\rceil}\leq\gamma_0^{(\mathrm{ar})}+\gamma_0^{(\mathrm{ar})} p\sum_{k=1}^\infty K^k\leq\gamma_0^{(\mathrm{ar})}+\frac{\gamma_0^{(\mathrm{ar})}pK}{1-K}.
$$

For an ARFIMA$(0,d,0)$, the autocorrelation function, $\rho_k^{(\textrm{afm})}$, is given in Equation$~\eqref{eqn:nfacr}$. The value of $\rho_k^{(\textrm{afm})}(d)$ is positive for $d>0$, negative for $d<0$ and zero for $d=0$.
Therefore, for $d\geq 0$, $\rho_k^{(\textrm{\textrm{afm}})}(d)\leq\rho_k^{(\textrm{afm})}(d^{\max})$
where $d^{\max}$ stands for the possible values of $d$ in the parametric space. For any stationary value of $d$, $\rho_k^{(\textrm{afm})}(d^{\max})\to 0$ as $k\to\infty$. For $d<0$, it is easy to see that $|d/(1-d)|$ is decreasing  while \linebreak $|(k-1+d)/(k-d)|$ for $k>1$ is increasing. Therefore, for $d<0$, $|\rho_k^{(\textrm{afm})}(d)|\leq\frac{1}{2}\prod_{i=2}^k\frac{i-1}{i}=\frac{1}{2k}.$
That is, both the lower and upper bounds for $\rho_k^{(\textrm{afm})}(d)$ converge to zero as $k\to\infty$.

To show the uniform convergence of $\rho_k$ in the ARFIMA$(p,d,q)$ model using the splitting method, it suffices to show the uniform convergence of the numerator in$~\eqref{eqn:Eq11.Cr}$, as the denominator is inferiorly bounded by $\sigma^2$ as a function of the parameters, regardless their values.

We now begin the proof for the ARFIMA$(p,d,0)$ model. The numerator in \linebreak Equation$~\eqref{eqn:Eq11.Cr}$ can be written as
\begin{equation} \label{eqn:P7.4}
\gamma_k=\sigma^{-2}\gamma_0^{(1)}\gamma_0^{(2)}\Big[\rho_0^{(1)}\rho_{-k}^{(2)}+\sum_{i=1}^\infty\rho_i^{(1)}\big(\rho_{i-k}^{(2)}+\rho_{i+k}^{(2)}\big)\Big],\quad k\geq 0.
\end{equation}
Now consider $\gamma_0^{(1)}$ and $\rho_0^{(1)}$ to be the autocovariance and autocorrelation functions of an ARFIMA($0,d,0$) process and $\gamma_0^{(2)}$ and $\rho_0^{(2)}$ to be the autocovariance and autocorrelation functions of an AR($p$) process. The value of $\gamma_0^{(1)}$ is given by $(-2d)!/(-d)!^2$. For $d\in(-1,1/2)$, $\gamma_0^{(1)}$ as a function of $d$ is well defined and continuous. As the parametric space is compact, $\gamma_0^{(1)}$ is clearly bounded. The value of $\gamma_0^{(2)}$ is given by $\gamma_0^{(2)}=\sigma^2/(1-\phi_1\rho_1-...-\phi_p\rho_p)$. Therefore, as the absolute values of the autocorrelations are smaller than 1, we obtain $\gamma_0^{(2)}\leq\frac{\sigma^2}{1-|\phi_1\rho_1+...+\phi_p\rho_p|}\leq\frac{\sigma^2}{1-K}.$

The expression inside the brackets in Equation$~\eqref{eqn:P7.4}$ has upper bound
\begin{equation} \label{eqn:UB}
\rho_0^{(1)}(d^{\max})K^{\lceil k/p\rceil}+\sum_{i=1}^\infty\rho_i^{(1)}(d^{\max})\left(K^{\lceil|i-k|/p\rceil}+K^{\lceil|i+k|/p\rceil}\right).
\end{equation}

In Equation$~\eqref{eqn:UB}$, if $k\to\infty$, clearly $\rho_0^{(1)}(d^{\max})K^{\lceil k/p\rceil}\to 0$. Additionally, \linebreak $\rho_i^{(1)}(d^{\max})\to 0$ as $i\to\infty$ and $K^{\lceil|i-k|/p\rceil}+K^{\lceil|i+k|/p\rceil}\to 0$ as $k\to\infty$. Furthermore, note that $\sum_{i=1}^\infty K^{\lceil|i+k|/p\rceil}\leq 2+\frac{2pK}{1-K}$ and $\sum_{i=1}^\infty K^{\lceil|i-k|/p\rceil}\leq 2+\frac{2pK}{1-K}$. Therefore, if we set $\rho_i^{(1)}=a_i$ and $K^{\lceil|i-k|/p\rceil}+K^{\lceil|i+k|/p\rceil}=w_{k,i}$, the criterias of Lemma 1 are satisfied, proving the uniform convergence to zero of the autocovariance function of the ARFIMA($p,d,0$) model.

The generalization of this conclusion to a general ARFIMA($p,d,q$) model is easy using the splitting method combining the autocovariance of the ARFIMA($p,q,0$) model and the MA$(q)$ model, both of which are now known to converge uniformly to zero.

\section{Proof of Theorem 1}

We preface the proof by the following lemma, which is used in the sequel.

\textbf{Lemma 2}: Define the function  $f:\mathbb{R}^m\to\mathbb{R}$, $f(a,b)=(a-b)'W(a-b)$, where $W$ is any positive definite matrix. Then, for any three vectors $a,b,c$, with the same dimension, $f(a,b)/2\leq f(a,c)+f(b,c)$.

\textit{Proof:} It is already known that $\sqrt{f(a,b)}\leq\sqrt{f(a,c)}+\sqrt{f(b,c)}$. Taking the square of the previous inequality and using the fact that $\sqrt{f(a,c)f(b,c)}\leq f(a,c)+f(b,c)$, leads to the required result.
$\square$

We proceed now to the proof of the theorem. The BCMDE searches for the $\lambda$ that minimizes $S(\lambda)$, given in Equation$~\eqref{eqn:Eq13}$. We will divide the proof in four parts.

\textbf{Part 1}: Let $\lambda_0$ be the true parameter value, then $f(\hat{\varrho},\varrho_T(\lambda_0))\xrightarrow{P} 0$ as $T\to\infty$, where $\xrightarrow{P}$ stands for convergence in probability.

Define the vector $\beta_T(\lambda)=\varrho_T(\lambda)-\varrho(\lambda)$. From Proposition 4, $\varrho_{T}(\lambda)$ converges uniformly to $\varrho(\lambda)$, so $\beta_T(\lambda)$ converges uniformly to zero as $T\to\infty$.
Additionally,
$f(\hat{\varrho},\varrho_T(\lambda_0))=(\hat{\varrho}-\varrho(\lambda_0))'W(\hat{\varrho}-\varrho(\lambda_0))-(\hat{\varrho}-\varrho(\lambda_0))'W\beta_T(\lambda_0)
-\beta_T(\lambda_0)W(\hat{\varrho}-\varrho(\lambda_0))+\beta_T(\lambda_0)'W\beta_T(\lambda_0).$

Due to the fact that $\hat{\varrho}-\varrho(\lambda_0)\to 0$ in probability as $T\to \infty$, so does $(\hat{\varrho}-\varrho(\lambda_0))'W(\hat{\varrho}-\varrho(\lambda_0))$, $(\hat{\varrho}-\varrho(\lambda_0))'W\beta_T(\lambda_0)$ and $\beta_T(\lambda_0)W(\hat{\varrho}-\varrho(\lambda_0))$, which implies that $\beta_T(\lambda_0)'W\beta_T(\lambda_0)$ also converges to zero.

\textbf{Part 2}: For $\lambda\neq \lambda_0$, $f(\hat{\varrho},\varrho_T(\lambda))\xrightarrow{P}c$, $c>0$, as $T\to\infty$.

We have already seen that, if $T\to\infty$, then $\beta_T(\lambda)\to 0$ and $\hat{\varrho}-\varrho(\lambda)$ converges in probability to a non-zero vector for $\lambda\neq\lambda_0$ (if the injectivity assumption of $\varrho$ is satisfied). Therefore, $\hat{\varrho}-\varrho(\lambda)-\beta_T(\lambda)$ converges in probability to a non-zero vector and $f(\hat{\varrho},\varrho_T(\lambda))=(\hat{\varrho}-\varrho_T(\lambda))'W(\hat{\varrho}-\varrho_T(\lambda))\to c$, $c>0$, using the assumption that $W$ is positive definite.

\textbf{Part 3}: For any neighborhood $V(\lambda_0)$ around $\lambda_0$, there exists a $L_2>0$ such that $f(\varrho_T(\lambda_0),\varrho_T(\lambda))\geq L_2$ if $\lambda\notin V(\lambda_0)$, for $T$ large enough.

In this case,
$f(\varrho_T(\lambda_0),\varrho_T(\lambda))=(\varrho(\lambda_0)-\varrho(\lambda))'W(\varrho(\lambda_0)-\varrho(\lambda))
+(\varrho(\lambda_0)-\varrho(\lambda))'W(\beta_T(\lambda_0)-\beta_T(\lambda))
+(\beta_T(\lambda_0)-\beta_T(\lambda))'W(\varrho(\lambda_0)-\varrho(\lambda))
+(\beta_T(\lambda_0)-\beta_T(\lambda))'W(\beta_T(\lambda_0)-\beta_T(\lambda)).$

As $\varrho(\lambda)$ is a continuous injective function and $\Lambda$ is a compact set, than there exists a $L>0$ such that, for $\lambda\notin V(\lambda_0)$, $(\varrho(\lambda_0)-\varrho(\lambda))'W(\varrho(\lambda_0)-\varrho(\lambda))>L$. Furthermore, because $\beta_T(\lambda)$ converges uniformly to zero as $T\to\infty$, then $(\varrho(\lambda_0)-\varrho(\lambda))'W(\beta_T(\lambda_0)-\beta_T(\lambda))\to 0$, $(\beta_T(\lambda_0)-\beta_T(\lambda))'W(\varrho(\lambda_0)-\varrho(\lambda))\to 0$ and $(\beta_T(\lambda_0)-\beta_T(\lambda))'W(\beta_T(\lambda_0)-\beta_T(\lambda))\to 0$, all of them uniformly. Therefore, for $T$ large enough, $f(\varrho_T(\lambda_0),\varrho_T(\lambda))\geq L_2$, for $L_2$ satisfying $0<L_2<L$.

\textbf{Part 4}: For any neighborhood $V(\lambda_0)$ of $\lambda_0$, $P(\hat{\lambda}\in V(\lambda_0))\to 1$.

In Part 1 it was shown that $f(\hat{\varrho},\varrho_T(\lambda_0))\xrightarrow{P} 0$. Therefore for any $L_2>0$, $P(f(\hat{\varrho},\varrho_T(d_0))<L_2/4)\to 1$. By Lemma 2, $f(\hat{\varrho},\varrho_T(\lambda))\geq f(\varrho_T(\lambda),\varrho_T(\lambda_0))/2-f(\hat{\varrho},\varrho_T(\lambda_0))$.
For $T$ large enough and $\lambda\notin V(\lambda_0)$, the first term in the right hand side of the above equation is greater or equal to $L_2/2$, while the probability that the second term is less than $L_2/4$ goes to 1. When both conditions are satisfied, $f(\hat{\varrho},\varrho_T(\lambda))\geq L_2/4$. In other words, the probability that the minimum of $f$ can be found at $V(\lambda_0)$, instead of outside $V(\lambda_0)$, goes to 1. The proof of the convergence in probability is completed. $\square$

\section{Proof of Proposition 5}
The value of $B_{T,k}^\rho$ under the conditions stated in proposition 5 is:
\begin{align}
\begin{split} \label{eqn:profprop6}
B_{T,k}^\rho=-\frac{2\sum_{t=1}^{T-k}\sum_{i=1}^T\rho_{|t-i|}}{(T-k)T}-\frac{24\sum_{t=1}^{T-k}\sum_{i=1}^T\tilde{z}_{t+k}\tilde{z}_i\rho_{|t-i|}}{(T-k)(T^3-T)}\\
+\frac{12[(T-k)^3-(T-k)(3k^2+1)]\sum_{i=1}^T\sum_{j=1}^T\tilde{z}_i\tilde{z}_j\rho_{|i-j|}}{(T-k)(T^3-T)^2}+\frac{\sum_{i=1}^T\sum_{j=1}^T\rho_{|i-j|}}{T^2}.\\
\end{split}
\end{align}

Note that $\sum_{i=1}^T\sum_{j=1}^T|\tilde{z}_i\tilde{z}_j|=(\sum_{i=1}^T|\tilde{z}_i|)^2$. With some algebra and using the fact that $\sum_{t=1}^Tt=T(T+1)/2$ we find that for even $T$, $\sum_{i=1}^T\sum_{j=1}^T|\tilde{z}_i\tilde{z}_j|=T^2/4$, while for odd $T$, $\sum_{i=1}^T\sum_{j=1}^T|\tilde{z}_i\tilde{z}_j|=(T^2-1)/4$. Therefore, we find that
$$\sum_{i=1}^T|w_{T,k,i}|\leq \frac{2T^2}{(T-k)T}+\frac{24T^4}{2(T-k)(T^3-T)}$$
$$+\frac{12[(T-k)^3-(T-k)(3k^2+1)]T^4}{2(T-k)(T^3-T)^2}+1,$$
which is an upper bound that converges to a constant as $T\to\infty$.

To check if $w_{T,k,i}\to 0$ as $T\to\infty$ note that $\sum_{i=1}^T\sum_{j=1}^T\rho_{|i-j|}=T\rho_0+\sum_{i=1}^{T-1}2(T-i)\rho_i$ and that
$$\sum_{i=1}^T\sum_{j=1}^T\tilde{z}_i\tilde{z}_j\rho_{|i-j|}=\rho_0\sum_{i=1}^T(t-\bar{t})^2+2\sum_{i=1}^{T=1}\rho_i\sum_{t=1}^{T-i}(t-\bar{t})(t+i-\bar{t}),$$
which is an expression of order $T^3$. In the above equation, $\bar{t}=\sum_{i=1}^Tt=t(t+1)/2.$ The expression $\sum_{i=1}^T\sum_{j=1}^T\tilde{z}_i\tilde{z}_j\rho_{|i-j|}$ is more complicated but it is also of order $T^3$. Indeed, all numerators in the right hand side of Equation \eqref{eqn:profprop6} are of a lower order of $T$ than the denominators, if we consider the individual weights. This proves that $w_{T,k,i}\to 0$ as $T\to\infty$.

\section{Proof of Theorem 2}

To evaluate the asymptotic distribution of the BCMDE, we should guarantee that the conditions of Theorem 3.2 of Newey and McFadden (1994) are satisfied. Define $\hat{g}_T(\lambda)=\hat{\varrho}-\varrho_T(\lambda)$ and let $\lambda_0$ be an interior point of $\Lambda$. We need to fulfill the following conditions:

i. $\hat{g}_T(\lambda)$ is continuously differentiable in a neighborhood  $V(\lambda_0)$ of $\lambda_0$;

ii. $\sqrt{T}\hat{g}_T(\lambda_0)\xrightarrow{\mathcal{D}} N(0,\Omega)$, $\Omega$ being any covariance matrix;

iii. There exists a $G(\lambda)$ that is continuos at $\lambda_0$ and $\sup_{\lambda\in V(\lambda_0)}||\nabla_\lambda\hat{g}_T(\lambda)-G(\lambda)||\to 0;$

iv. For $G=G(\lambda_0)$, $G'WG$ is non-singular.

\noindent If all these requirements are satisfied, then $\sqrt{T}(\hat{\lambda}-\lambda)\xrightarrow{\mathcal{D}} N(0,\Omega)$.

We proceed now to verify if the above conditions are valid in the case of an ARFIMA ($0,d,0$).

\textbf{Item i}. In this case the vector $\varrho_T(d)$ is given by $\varrho_T(d)=(\rho_{T,k_1},...,\rho_{T,k_m})$ where $\rho_{T,k}=\frac{\rho_k+B_{T,k}^\rho}{1+B_{T,0}^\rho}$. That is, $\rho_{T,k}$ is the ratio between two linear combinations of autocorrelations. It is already known that the theoretical autocorrelations of an ARFIMA$(0,d,0)$ model are continuously differentiable. Therefore, both the numerator and the denominator of $\rho_{T,k}$ are continuously differentiable. Moreover the denominator does not vanish because it is the expectation of a non-negative random variable. Thus, as the derivative of $\hat{\rho}$ is equal to zero, $\hat{g}_T(d)$ is continuously differentiable in a neighborhood $V(\lambda_0)$ of $d$.

\textbf{Item ii}. Note first that
\begin{equation} \label{eqn:conv.vic}
\sqrt{T}\hat{g}_T(d_0)=\sqrt{T}(\hat{\varrho}-\varrho(d_0))-\sqrt{T}\beta_T(d_0).
\end{equation}
The first term in the right hand side of Equation$~\eqref{eqn:conv.vic}$ converges to $N(0,C)$ when $d<0.25$ as shown in Hosking (1996), $C$ being the asymptotic covariance matrix of $\hat{\varrho}$. For the second term, $\beta_T(d_0)$, we will first write each element of the vector as
\begin{equation} \label{eqn:conv.vic2}
\sqrt{T}\beta_{T,k}(d_0)=\frac{\rho_k(d_0)\sqrt{T}B_{T,0}^\rho-\sqrt{T}B_{T,k}^\rho}{1+B_{T,0}^\rho}.
\end{equation}
We have already seen that the denominator in$~\eqref{eqn:conv.vic2}$ converges to one. Regarding the numerator, $B_{T,0}^\rho$ and $B_{T,k}^\rho$ stand for the bias of the autocovariance (for $\gamma_0=1$) for lags 0 and $k$, respectively. Hosking (1996) shows that the bias of the sample autocovariance decays at the rate $T^{2d-1}$. Therefore, for $d<0.25$, $\rho_k(d_0)\sqrt{T}B_{T,0}^\rho-\sqrt{T}B_{T,k}^\rho\to 0$.

\textbf{Item iii}. The derivative of $\hat{g}_T(d)$ is given by $\nabla_d\hat{g}_T(d)=0-\nabla_d\varrho_T(d)$. The elements of vector $\varrho_T(d)$ are approximations of the expectancies of $\hat{\varrho}$ given in$~\eqref{eqn:Ro}$. Therefore,
\begin{align*}
\nabla_d\rho_{T,k}(d)&=\frac{[\nabla_d\rho_k(d)-\nabla_dB_{T,k}^\rho(d)][1-B_{T,0}^\rho(d)]}{[1-B_{T,0}^\rho(d)]^2}\\
&-\frac{[\rho_k(d)-B_{T,k}^\rho(d)][-\nabla_dB_{T,0}^\rho(d)]}{[1-B_{T,0}^\rho(d)]^2}.\\
\end{align*}
It has already been shown in the proof of Proposition 4 that $B_{T,k}^\rho(d)$ converges uniformly to 0. Thus we only need to analyze the behavior of the derivatives of $B_{T,k}(d)$.

First of all, we investigate the behavior of $\nabla_d\rho_k(d)$. We can write $\nabla_d\rho_k(d)$ as $\nabla_d\rho_k(d)=\nabla_d(\log |\rho_k(d)|)\rho_k(d)$ for any $d\ne 0$, where $\rho_k(d)$ is given in Equation$~\eqref{eqn:Eq11.Cr}$. Thus,
$$\nabla_d\rho_k(d)=\frac{1}{(1-d)^2}\prod_{i=2}^k\frac{i-1+d}{i-d}.$$
At $d=0$, $\nabla_d\rho_k(d)=1/k$, which obviously converges to zero as $k\to\infty$. For $d\ne 0$ note that $$\log|\rho_k(d)|=\log|d|+\sum_{i=2}^k\log(i-1+d)-\sum_{i=1}^k\log(i-d).$$
Consequently,  $$\nabla_d(\log|\rho_k(d)|)=\frac{\textrm{sgn}(d)}{|d|}+\sum_{i=2}^k\frac{1}{i-1+d}+\sum_{i=1}^k\frac{1}{i-d},$$
where $\textrm{sgn}(d)$ is the signal of $d$. Therefore, $$\nabla_d\rho_k(d)=\frac{1}{1-d}\prod_{i=2}^k\frac{i-1+d}{i-d}+\left(\sum_{i=2}^k\frac{1}{i-1+d}+\sum_{i=1}^k\frac{1}{i-d}\right)\prod_{i=1}^k\frac{i-1+d}{i-d}.$$

Let $d_{\textrm{sup}}$ and $d_{\textrm{inf}}$ be, respectively, the supremum and infimum of $\Lambda$, the parameter space. Then, an upper bound for the absolute value of $\nabla_d\rho_k(d)$ is given by
\begin{align*}
\textrm{UB}(\nabla_d\rho_k(d))&=\frac{1}{1-d_{sup}}\prod_{i=2}^k\frac{i-1+d_{sup}}{i-d_{sup}}\\
&+\left(\sum_{i=2}^k\frac{1}{i-1+d_{inf}}+\sum_{i=1}^k\frac{1}{i-d_{sup}}\right)\prod_{i=1}^k\frac{i-1+d_{sup}}{i-d_{sup}}.\\
\end{align*}
Note that $\prod_{i=2}^k\frac{i-1+d_{sup}}{i-d_{sup}}=O(k^{2d_{sup}-1})$, $\sum_{i=2}^k\frac{1}{i-1+d_{inf}}=O(\log(k))$, \linebreak$\sum_{i=1}^k\frac{1}{i-d_{sup}}=O(\log(k))$ and $\prod_{i=1}^k\frac{i-1+d_{sup}}{i-d_{sup}}=O(k^{2d_{sup}-1})$, therefore \linebreak $\textrm{UB}(\nabla_d\rho_k(d))\to 0$ as $k\to\infty$. That is, $\nabla_d\rho_k(d)$ converges uniformly to zero and so does $\nabla_dB_{T,k}(d)$, as it is a linear combination of $\nabla_d\rho_1(d),...,\nabla_d\rho_{T-1}(d)$ that satisfies the properties of Lemma 1.

\textbf{Item iv} As the matrix $W$ is by definition positive definite, it suffices to show that at least one element of $G(d_0)=\nabla_d\rho_{k}(d_0)$ is nonzero. One way to guarantee that is to use the first lag, $\rho_1(d)$, in vector $\varrho$ as $\nabla_d\rho_1(d)>0$ for all values of $d$ in the invertibility and stationarity regions.
$\square$

\FloatBarrier


\section*{References}

\bibliography{sample}


\leftskip 0.2in
\parindent -0.2in
Andel, J. (1986). Long memory time series models. {\it Kybernetika} 22, 105-123.

Arnau, J. \& Bono, R. (2001). Autocorrelation and bias in short time-series: an alternative estimator. {\it Quality \& Quantity} 35, 365-387.

Bertelli, S. \& Caporin, M. (2002). A note on calculating autocovariances of long memory processes. {\it Journal of Time Series Analysis} 23, 503-508.

Box, G. E. P. \& Jenkins, G. M. (1976). {\it Time Series Analysis: Forecasting and Control}, 2nd ed. San Francisco: Holden Day.

Brockwell, P. J. \& Davis, R. A. (1991). {\it Time series: Theory and Methods}, 2nd ed. New York: Springer-Verlag.

Fox, R. \& Taqqu, M. S. (1986). Large-sample properties of parameter estimates for strongly dependent stationary Gaussian time series. {\it The Annals of Statistics} 14, 2, 517-532.

Geweke, J. \& Porter-Hudak, S. (1983). The estimation and application of long memory time series model. {\it Journal of Time Series Analysis} 4, 221-238.

Granger, C.M.G. \& Joyeux, R. (1980). An introduction to long memory time series models and fractional differencing. {\it Journal of Time Series Analysis} 1, 15-29.

Haslett, J. \& Raftery, A. E. (1989). Space-time modelling with long-memory dependence: Assessing Ireland's wind power resource. {\it Journal of Applied Statistics} 38, 1-50.

Hassani, H., Leonenko, N. \& Patterson, K. (2012). The sample autocorrelation function and the detection of long-memory processes. {\it Physica A} 391, 6367-6379.

Hosking, J. R. M. (1981). Fractional Differencing. {\it Biometrika} 68, 1, 165-176.

Hosking, J. R. M. (1996). Asymptotic distribution of the sample mean, autocovariance and autocorrelation of long-memory time series. {\it Journal of Econometrics} 73, 261-284.

Huitema, B. E. \& McKean, J. W. (1994). Two reduced-bias autocorrelation estimator: $r_{F1}$ and $r_{F2}$. {\it Perceptual and Motor Skills} 78, 323-330.

Mayoral, L. (2007). Minimum distance estimation of stationary and non-stationary ARFIMA processes. {\it Econometrics Journal} 10, 124-148.

Palma, W. (2007). {\it Long-Memory Time Series}. Hoboken, New Jersey: Wiley.

Priestley, M. B. (1981). {\it Spectral Analysis and Time Series}. London: Elsevier.

Rea, W., Oxley, L., Reale, M. \& Brown, J. (2013). Not all estimators are born equal: The empirical properties of some estimators of long memory. {\it Mathematics and Computers in Simulation} 93, 29-42.

Reisen, V. A. (1994). Estimation of the fractional difference parameter in the ARIMA$(p,d,q)$ model using the smoothed periodogram. {\it Journal of Time Series Analysis} 15, 335-350.

Tieslau, M. A., Schmidt, P. \& Baillie, R. T. (1996). A minimum distance estimator for long-memory processes. {\it Journal of Econometrics} 71, 249-264.

Zevallos, M. \& Palma, W. (2013). Minimum distance estimation of ARFIMA processes. {\it Computational Statistics and Data Analysis} 58, 242-256.

Whittle, P. (1951). {\it Hypothesis Testing in Time Series Analysis}. New York: Hafner.


\end{document}